\documentstyle[11pt,aaspp4]{article}

\slugcomment{accepted for the publication in the Astrophysical Journal, 
{\it Letters}}

\def\ros{{\it ROSAT }}
\def\iras{{\it IRAS }}

\newcommand{\tqkev}{$3\over4$~keV}

\newcommand{\hi}{H{\small ~I}}
\newcommand{\hii}{H{\small ~II}}

\def\fdg{\hbox{$.\!\!^\circ$}}
\def\farcm{\hbox{$.\mkern-4mu^\prime$}}

\def\la{\mathrel{\hbox{\rlap{\hbox{\lower4pt\hbox{$\sim$}}}\hbox{$<$}}}}
\def\ga{\mathrel{\hbox{\rlap{\hbox{\lower4pt\hbox{$\sim$}}}\hbox{$>$}}}}

\lefthead{Park, Finley, Snowden, \& Dame}
\righthead{X-ray Shadows}

\begin{document}
\title{EVIDENCE FOR AN X--RAY EMITTING GALACTIC BULGE:
SHADOWS CAST BY DISTANT MOLECULAR GAS }
\author{\bf Sangwook Park and John P. Finley}
\affil{Department of Physics,  Purdue University \\
 1396 Physics Building, West Lafayette, IN. 47907 \\
Electronic Mail: (parksan,finley)@purds1.physics.purdue.edu}
\author{\bf S. L. Snowden\altaffilmark{1}}
\affil{NASA/Goddard Space Flight Center, Code 662, Greenbelt, MD. 20771 \\
Electronic Mail: snowden@lheavx.gsfc.nasa.gov}
\author{\bf and \\
           T. M. Dame}
\affil{Harvard-Smithsonian Center for Astrophysics, 60 Garden Street,
Cambridge, MA. 02138 \\
Electronic Mail: tdame@cfa.harvard.edu}

\altaffiltext{1}{Universities Space Research Association}

\begin{abstract}
A mosaic of 7 \ros PSPC pointed observations in the direction of
($l,b\sim10^\circ,0^{\circ}$) reveals deep X-ray shadows in the 
$0.5-2.0$~keV band cast by dense molecular gas. The comparison
between the observed on-cloud and off-cloud X-ray fluxes indicates that
$\sim43$\% of the diffuse X-ray background in this direction in both the 
\tqkev\ and 1.5~keV bands originates behind the molecular gas, 
which is located at 2--4 kpc from the Sun.  Given the short mean free 
path of X-rays in the \tqkev\ band in the Galactic plane ($\sim1$~kpc 
assuming an average space density of 1~cm$^{-3}$), this large percentage 
of the observed flux which originates beyond the molecular gas most 
likely indicates a strong enhancement in the distribution of X-ray emitting 
gas in the Galactic center region, possibly associated with a Galactic 
X-ray bulge.  

\end{abstract}


\section {INTRODUCTION}

The origin of the $\sim0.5-2.0$~keV diffuse X-ray background has long eluded 
a satisfactory explanation. The extrapolation of the extragalactic power law 
($I_{X}\sim11 E^{-1.4}$ photons cm$^{-2}$ s$^{-1}$ sr$^{-1}$ keV$^{-1}$) from 
the $2-10$~keV band (\cite{hea71}; \cite{bea79}) accounts for $\la50$\% 
of the observed X-ray intensity at high Galactic latitudes in the 
$\sim0.5-1.0$~keV band (\cite{mea83}).  
While there remains some discussion as to the exact normalization of the 
power law and shape of the spectrum below 2~keV (e.g., 9.2$E^{-1.4}$ and no 
appreciable excess between 1 and 2~keV, \cite{gen95}; 10.8$E^{-1.4}$, 
\cite{cfg96}), all measurements agree that there is a strong excess 
above extrapolation of the 2 -- 10 keV power law spectrum 
in the $0.5-1.0$~keV band.  An upper limit to the 
contribution from the Local Hot Bubble (LHB, \cite{sea90}) which 
extends out from 50 to a couple hundred parsecs is 
$\sim20$\% of the nominal high-latitude intensity, as revealed by a 
high-latitude X-ray shadow due to a nearby molecular cloud MBM 12 
(\cite{smv93}). The contributions from stellar sources (e.g., late type 
stars) have been considered to explain the observed excess X-ray intensity 
in the Galactic plane (\cite{rea81}; \cite{ss90}; \cite{w92}; \cite{os92}), 
however they do not produce a sufficient flux. 
Although a substantial fraction of the
thermal excess of the diffuse X-ray background at high Galactic
latitudes is from discrete extragalactic sources (e.g., AGNs),
at least 25\% of the diffuse X-ray emission is not due to AGNs 
(\cite{hea93}). A significant portion of the diffuse X-rays could still 
be Galactic, yet from beyond the LHB even at high latitudes. 
The problem is even more severe when the 
intensity near the Galactic plane is considered.  The Galactic plane is 
optically thick at these energies so {\it all} of the observed flux 
{\it must} be Galactic in origin.  

The Galactic diffuse X-ray emission beyond the LHB 
strongly suggests an abundance of Galactic gas at 
$\sim3\times10^6$~K (\cite{sea82}).  The distribution of the X-ray
emitting material in the Galactic plane beyond obvious supernova remnants 
and stellar wind bubbles is, however, poorly known. Snowden et al.
(1996) have recently suggested that the strong X-ray intensity seen
above and below the plane in the general direction of the Galactic
center is due to a Galactic X-ray bulge.  The
best way to study the Galactic distribution of the X-ray emitting gas is to 
search for shadows in the diffuse X-ray background cast by absorbing high 
column-density molecular gas for which the distance can be determined.  
The utility of this 
approach has been demonstrated in the literature in the case of nearby
neutral and molecular clouds at high Galactic latitudes (e.g., \cite{bm91}; 
\cite{sea91}; \cite{smv93}; \cite{wy95}; \cite{ksv96}). Here we report the 
detection of 
deep X-ray shadows cast by dense molecular gas in the inner Galaxy.  These 
shadows imply an extensive flux of diffuse X-rays from the Galactic center 
region.

The line of sight of our study region, ($l,b\sim10^\circ,0^\circ$), 
extensively samples the inner Galaxy (1.5~kpc minimum radius) but still 
allows reliable velocity-derived distances for the molecular gas.  
This region has been covered by a mosaic of 7 independent \ros PSPC pointed 
observations, and by the Galactic CO survey of Bitran (1987).  We use the CO 
to derive both kinematic distances and molecular column densities for the 
absorbing gas, and the \iras 100 $\mu$m intensity as a tracer of total gas
column density.  

The data used for this study are described
in \S\ref{sec:data}.  The analysis and the implications of the results are
discussed in \S\ref{sec:analysis} and a summary and some conclusions 
are presented in \S\ref{sec:conclusions}.

\section{\label{sec:data} DATA}
The CO data used in the present study were taken from the Galactic
survey of Bitran (1987), which was carried out with the 1.2 m
millimeter-wave telescope at Cerro Tololo, Chile. At 115 GHz, the
frequency of the J = 1 -- 0 rotational transition, this
telescope has a beamwidth of $8\farcm8$ (FWHM) and its 256-channel
spectrometer provides a velocity resolution of 1.3 km s$^{-1}$ and total
bandwidth of 333 km s$^{-1}$. Observations were spaced roughly every
beamwidth ($7\farcm5$) on a Galactic grid, and the rms sensitivity of 
0.14~K (T$_{mb}$) was more than adequate for the present purpose.

The \iras 100 $\mu$m data (\cite{wea94}) for this region were obtained
through the High Energy Astrophysics Science Archive (HEASARC) SkyView
facility which is maintained by NASA/GSFC. The displayed \iras 100 $\mu$m
field (Figure~\ref{fig:irxr}) is $4\fdg3\times4\fdg3$ centered at 
($l,b=9.84^\circ,-0.15^\circ$). The infrared intensity for this field 
ranges from $\sim200$~MJy~sr$^{-1}$ to $10^4$~MJy~sr$^{-1}$ and the
angular resolution is $\sim5'$.

The 7 \ros (\cite{tru92}) PSPC pointed observations are listed 
in Table~\ref{tbl:observations}.  Data from all 7 pointings 
were obtained through the HEASARC \ros public archive.  Since non-cosmic 
contamination is not negligible in the study of the diffuse X-ray background, 
all identified non-cosmic contributions to the counting rate must be handled 
thoroughly.  The particle background, scattered solar X-ray background, 
and long-term enhancements are modeled and subtracted from the data by 
following the methods described in the literature (\cite{smbm94}, 
and references therein). Contamination due to short-term enhancements including
auroral X-rays, solar flares, and enhanced charged-particle rates encountered
near the South Atlantic Anomaly and particle belts are removed by excluding all
observation time intervals which display anomalous peaks in their light curves.
For these 7 pointings, the modeled non-cosmic contamination ranges up to
$\sim50$\% of the total observed X-ray counts and the mean contribution is
$\sim20$\%.

The individual PSPC pointings with all identified non-cosmic contamination,
point sources, and discrete small-scale extended emission features (e.g.,
G11.2--0.3 and W30) removed are merged into large-area mosaics in two bands:
the $0.44-1.21$~keV band (\tqkev) and the $0.73-2.04$~keV band (1.5~keV).  The
large amount of spectral overlap between the two bands is due to the relatively
poor spectral resolution of the proportional counter.  The determination 
of the relative offsets in the zero level between the individual
observations is performed by comparing the average count rates in the
overlapping regions between all pairs of observations. The contribution
from this correction is typically small and is $\la3$\% of the
total counts for this field of view.  The software for this
task (\cite{s94}) was provided by the US \ros Science Data Center (USRSDC) 
at NASA/GSFC.  The feasibility of the software has been successfully 
demonstrated by a mosaic of the Large Magellanic Cloud (\cite{sp94}).

The final \ros PSPC mosaics of the region ($l,b\sim10^\circ,0^\circ$) 
are displayed in Figures~\ref{fig:irxr}a and 
\ref{fig:irxr}b.  For purposes of display, the data were binned 
into $5'$ pixels and smoothed. For the analysis of this paper, the
unsmoothed $5'$ pixels were used, usually with further binning. This pixel 
size was selected since it is reasonably comparable to the 
spatial resolution of the \iras 100~$\mu$m and the CO data without
losing most of the detailed spatial structure of the X-ray emission.  
Combined as a mosaic, the average exposure for this field is $\sim18$~ks 
yielding $\sim15$\% statistical errors for individual pixels.

\section{\label{sec:analysis} ANALYSIS AND DISCUSSION}

The Galactic plane in the range {\it l}
$\sim$ 10$^{\circ}$ -- 11$^{\circ}$ shows a broad maximum in the far
infrared and a corresponding minimum in 0.5 -- 2.0 keV band X-rays 
(Figure~\ref{fig:irxr}a and Figure~\ref{fig:irxr}b). 
Although the intense point-like features in the
far infrared map (Figure~\ref{fig:irxr}) arise from well-known Galactic
\hii~regions such as W31 (the cluster of sources near the center of the
map), the more extended emission is considered to be a fairly reliable
tracer of total gas column density on a Galactic scale (\cite{bdt90}).
No significant CO concentration is associated with any of the \iras
cores in all velocity intervals except for the one associated
with SGR 1806--20 at 15 kpc (Figure~\ref{fig:co}a, \cite{cea96}). 
Since the in-plane gas density peaks about half-way between the Sun and
the Galactic center (\cite{d93}), the anticorrelation seen in 
Figure~\ref{fig:irxr} suggests that a significant fraction of the X-rays
arise from beyond the bulk of the gas, most likely in the Galactic
bulge.

This location of the bulk of the X-ray emitting gas is supported by the 
more detailed
anticorrelation seen between X-rays and the dense molecular component of
the gas as traced by CO. In Figure~\ref{fig:co}, the CO emission is
partitioned into 4 equal intervals of velocity and overlaid as contours
on the 1.5 keV band X-ray map that has been smoothed to match the
angular resolution of the CO data. These four maps account for nearly
all of the CO emission in this direction; similar CO maps at higher
velocities are blank or nearly so. The CO emission is strongest and most
widespread in the velocity range 15 -- 30 km s$^{-1}$
(Figure~\ref{fig:co}b), and it is this emission which shows the best 
detailed anticorrelation with the X-rays. According to the Galactic
rotation curve of Burton (1988), such emission must arise either from
the near distance interval 2.5 -- 3.9 kpc (as indicated above the plot)
or from the corresponding far interval 12.8 -- 14.3 kpc. With such a
large difference between the near and far distances, CO maps at the
scale and resolution of Figure~\ref{fig:co} would tend to be dominated
by near side material (given the typical scale height), and the 
anticorrelation with the X-rays implies
that such is probably the case here.

Figure~\ref{fig:co}b displays apparent X-ray shadows cast by 
molecular gas at different velocities and distances. The 
Y-shaped shadow extending above the plane near {\it l} $\sim$
10.5$^{\circ}$ is perhaps the most notable, both because of its detailed
match to the CO and the relatively high velocity ($\sim30$ km s$^{-1}$)
and distance (3.9 $\pm$ 0.7 kpc) of the absorbing gas; this same vertical
structure is even partially seen in the 30 -- 45 km s$^{-1}$ map
(Figure~\ref{fig:co}c). The clarity of this shadow might be due in
part to its relatively high latitude, where confusion by background
material at the far kinematic distance should be negligible. Two other
well-defined shadows are seen near {\it l} $\sim$ 8.6$^{\circ}$, and are 
also well displaced from the plane: one at {\it b} $\sim$
0.5$^{\circ}$ and the other at {\it b} $\sim$ --0.7$^{\circ}$. Both
features have gas velocities near 18 km s$^{-1}$, corresponding to a
near kinematic distance of 3.1 $\pm$ 0.5 kpc. The most notable exception
to the general anticorrelation seen in Figure~\ref{fig:co} is the
region near ({\it l, b} $\sim$ 11.25$^{\circ}$, --0.5$^{\circ}$), which
is very bright in X-rays yet shows relatively strong CO emission at
distances greater than 3.9 kpc (Figures~\ref{fig:co}c).  
The bulk of these X-rays most likely arise from a foreground diffuse X-ray 
emission feature. 

In the regions where the absorption is clear, 
($l,b\sim10.3^\circ,-0.2^\circ$), ($l,b\sim8.6^\circ,-0.7^\circ$), and 
($l,b\sim10.5^\circ,0.5^\circ$) (see
Figure~\ref{fig:co}b), the H$_2$ column density for the molecular gas
is estimated to be $\ga~1\times10^{22}$~cm$^{-2}$ by applying a CO-H$_2$
conversion factor $\frac{H_2}{CO}$ = 2 $\times$ 10$^{20}$ $\frac{cm^{-2}}{K km
s^{-1}}$. This large molecular column density
indicates that the molecular gas in these directions is optically thick for
the 0.5--2.0~keV X-rays (i.e. in the $3\over4$ keV band 1 optical depth
is $\sim$2.7$\times$10$^{21}$ cm$^{-2}$ and $\sim$4$\times$10$^{21}$
cm$^{-2}$ in the 1.5 keV band assuming the theoretical cross
section of Morrison \& McCammon (1983) and a 10$^{6.6}$K thermal plasma.)
It is therefore possible to
directly extract the on-cloud and off-cloud X-ray intensities in order to
estimate the foreground and distant fractions of the observed X-ray flux 
rather than fitting the standard two-component absorption model (e.g.,
\cite{smv93}). 

The on-cloud X-ray emission is derived from a 1200 arcmin$^{2}$ extraction
region centered on ($l,b\sim10.3^\circ,-0.2^\circ$), a 600 arcmin$^{2}$ 
extraction region centered on ($l,b\sim8.6^\circ,-0.7^\circ$), and a 300 
arcmin$^{2}$ extraction region centered on ($l,b\sim10.5^\circ,0.5^\circ$).
The range of on-cloud X-ray intensity in the \tqkev\ band from these 3 regions 
is 53 -- 57 $\times$ 10$^{-6}$~counts~s$^{-1}$~arcmin$^{-2}$, 
while the 1.5~keV band intensity is $\sim$
$94\times10^{-6}$~counts~s$^{-1}$~arcmin$^{-2}$ in all three regions. 
Several regions with I$_{CO}<28$~K km s$^{-1}$ (i.e., H$_{2}$ density
$\la$ 5 $\times$ 10$^{21}$ cm$^{-2}$) were selected to estimate the
off-cloud X-ray intensities in the field: a 750 arcmin$^{2}$ region 
centered on ($l,b\sim9.9^\circ,0.4^\circ$), a 400 arcmin$^{2}$ region
centered on ($l,b\sim11.3^\circ,0.5^\circ$), a 600 arcmin$^{2}$ region
centered on ($l,b\sim11.0^\circ,-1.1^\circ$), and a 375 arcmin$^{2}$ region 
centered on ($l,b\sim9.0^\circ,-1.3^\circ$).
The brightest regions (i.e.,
({\it l, b} $\sim$ 11.5$^{\circ}$, --0.5$^{\circ}$) and 
({\it l, b} $\sim$ 9.3$^{\circ}$, --1.0$^{\circ}$)) were
excluded to avoid contamination by possible discrete X-ray emission features. 
The off-cloud X-ray intensities, the sum of foreground and distant emission, 
lie in the range 
$\sim87-109\times10^{-6}$~counts~s$^{-1}$~arcmin$^{-2}$ in the \tqkev\ band and
$\sim149-179\times10^{-6}$~counts~s$^{-1}$~arcmin$^{-2}$ in the 1.5~keV band.
The observed on-cloud and average off-cloud X-ray intensities for this
field of view are summarized in Table~\ref{tbl:fluxes}. 
The off-cloud X-ray intensities are consistent with the range of 
$0.5-2.0$~keV \ros all-sky survey values (\cite{seasur95}) in the 
first quadrant of the Galactic plane.

The average on-cloud to off-cloud X-ray intensity ratio in the \tqkev\ band 
is $0.57\pm0.09$, which implies that $43\pm9$\% of the observed X-ray flux in 
this band originates behind the molecular cloud. Given 
the probable large X-ray absorption optical depth foreground to the cloud 
region ($\tau\sim3.4$, assuming an average space density of 
$\sim1$~\hi~cm$^{-3}$ in the midplane (\cite{sc92}) and a distance of 
$\sim3$~kpc) 
a bright background emission region is implied:
$\sim1229\times10^{-6}$~counts~s$^{-1}$~arcmin$^{-2}$ even assuming no 
additional  absorption between the cloud and the distant emission region.  
This value is roughly an order of magnitude greater than the nominal 
high-latitude intensity ($\sim130\times10^{-6}$~counts~s$^{-1}$~arcmin$^{-2}$). 
This emission is consistent with the extrapolation to the plane of the 
high X-ray intensities seen above and below the plane (interpreted as 
a Galactic X-ray bulge by Snowden et al. (1996)) in this direction.

The average on-cloud to off-cloud X-ray flux ratio in the 1.5~keV band of
$0.56\pm0.08$ is nearly the same as that for the \tqkev\ band, and the 
foreground optical depth is $\tau\sim2.3$ (assuming 1 optical depth
$\sim$4 $\times$ 10$^{21}$ cm$^{-2}$).  The interpretation is the same 
as for the \tqkev\ band shadow, that there is a strong enhancement in the 
filling factor of X-ray emitting gas in the Galactic plane beyond 
$\sim3$~kpc in the direction of the Galactic center.  The implied 
background emission, again with no additional absorption, is
$\sim725\times10^{-6}$~counts~s$^{-1}$~arcmin$^{-2}$.  The deabsorbed 
1.5~keV to \tqkev\ band ratio for the distant emission, $\sim0.59$, suggests 
a thermal emission temperature of $10^{6.7}$~K.  This is again consistent 
with the Galactic X-ray bulge of Snowden et al. (1996).

\section{\label{sec:conclusions} SUMMARY AND CONCLUSIONS}

We have presented and discussed the implications of the deep 
$0.5-2.0$~keV band X-ray shadows in the Galactic plane cast by dense 
inner-Galaxy molecular gas toward {\it l} $\sim$ 10$^{\circ}$. The 
striking result that half of the observed intensity originates beyond an 
optical depth of 2.3--3.4 requires that the emitting gas is non-uniformly 
distributed in the Galactic disk and has a strong enhancement near the Galactic 
center.  
Its existence is also consistent with the extrapolation to the plane of 
the emission from a probable Galactic X-ray bulge (\cite{seasur96}).

While some of the emission responsible for the foreground component 
could arise in the nearby ($\sim150$~pc with a $\sim100$~pc radius) 
Sco-Cen Bubble, a part of this emission in all likelihood originates 
from another source component or superposition of multiple components 
which may be distributed over the $\sim3$~kpc to the molecular gas.  
This additional component(s) presumably contributes as well to the observed 
nonzero flux along the Galactic plane observed in the anticenter direction 
where any extragalactic emission is still completely absorbed.  Further 
analysis of additional shadowing targets will help clarify the situation.

\acknowledgments

{The authors would like to thank Jay Lockman for helpful discussion and
direction and the referee Dan McCammon for his thoughtful review of
this paper.
This research has made use of data obtained through the High Energy
Astrophysics Science Archive Research Center Online Service, provided by
the NASA-Goddard Space Flight Center and was supported in part by NASA
grant NAG 5-2492 and the Purdue Research Foundation. }

\clearpage

\begin{deluxetable}{lcccc}
\footnotesize
\tablecaption{List of \ros PSPC Observations. 
              \label{tbl:observations}}
\tablewidth{0pt}
\tablehead{
\colhead{Observation ID} & \colhead{$l$} & \colhead{$b$}
& \colhead{Exposure (ks)} & \colhead{Date}
}
\startdata
WP500197N00 &  8.76 & --0.11 & 3.14 & 5 Oct 1992       \nl
WP200717    & 10.80 &  0.39 & 8.06 & $7-8$ Mar 1992   \nl
RP500150N00 & 11.12 & --0.35 & 4.87 & 4 Mar 1993       \nl
WP900145    &  9.96 & --0.24 & 2.44 & $18-27$ Mar 1991 \nl
WP900399N00 &  9.96 & --0.24 & 9.39 & $2-3$ Apr 1993   \nl
WP180034N00 &  9.71 & --0.39 & 1.38 & $9-10$ Oct 1993  \nl
RP201060N00 &  9.24 & --0.61 & 8.50 & $2-3$ Apr 1993   \nl
\enddata

\end{deluxetable}

\clearpage

\begin{deluxetable}{cccc}
\centering
\footnotesize
\tablecaption{The Observed On-Cloud and Off-Cloud X-ray Fluxes.
\label{tbl:fluxes}}
\tablewidth{0pt}
\tablehead{
\colhead{Band} & \colhead{On-Cloud\tablenotemark{a}} &
 \colhead{Off-Cloud\tablenotemark{a}} & \colhead{Ratio} 
}
\startdata
\tqkev\ & $55\pm6$  & $96\pm10$  & $0.57\pm0.09$ \nl
1.5 keV & $94\pm9$  & $168\pm17$ & $0.56\pm0.08$ \nl
\enddata

\tablenotetext{a}{On-cloud and off-cloud fluxes in units of
$10^{-6}$~counts~s$^{-1}$~arcmin$^{-2}$. The errors include both the
formal statistical uncertainty as well as a systematic contribution
related to the selection of the on-cloud and off-cloud regions.}

\end{deluxetable}

\clearpage

\figcaption[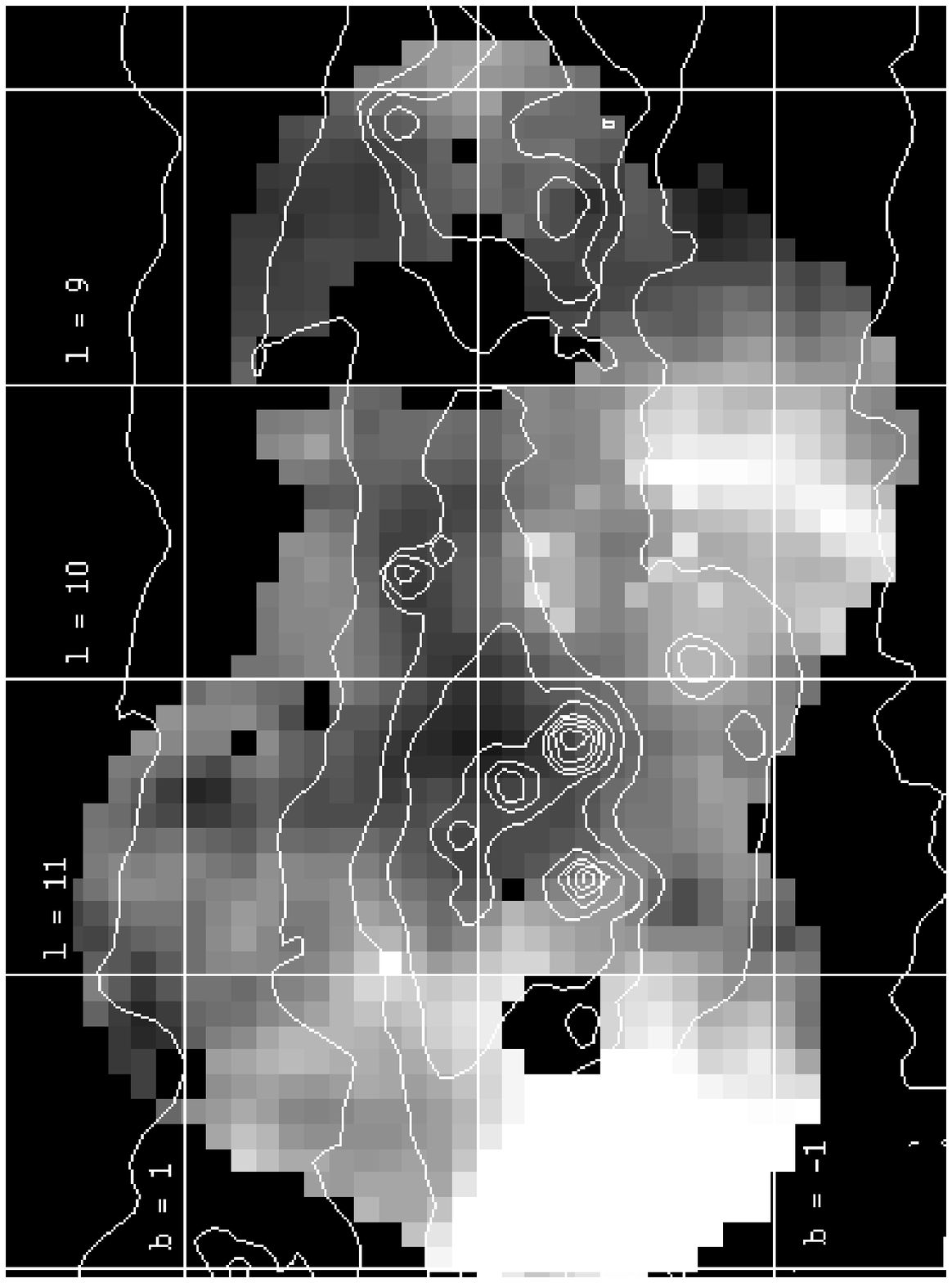, 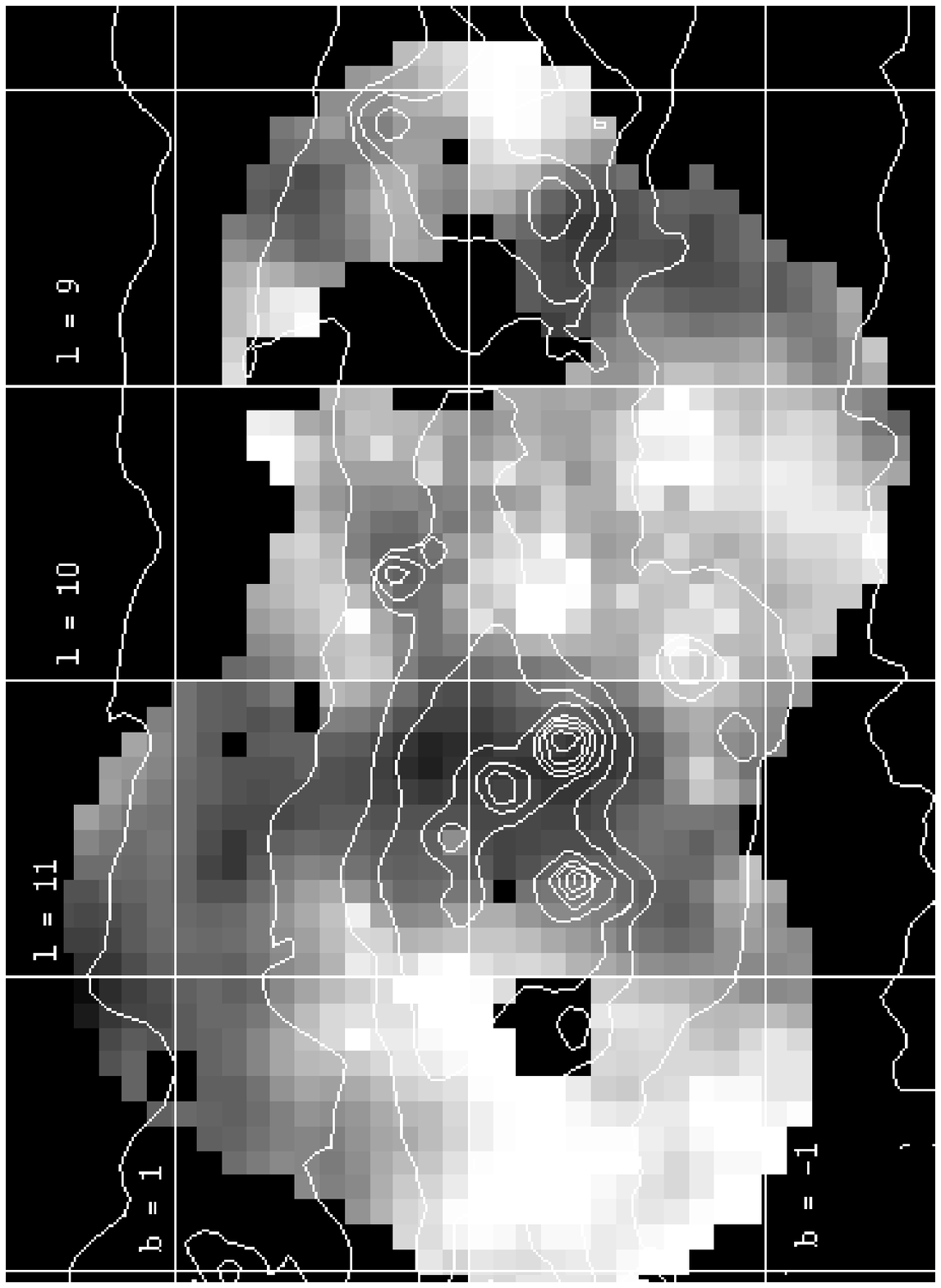]
{The X-ray images of the Galactic plane near $l\sim10^\circ$.
These cover an $\sim14$~deg$^{2}$ region between $l= 7.5^\circ$ 
and $l=12^\circ$ and $\left|b\right|\leq1.5^\circ$ including the W30 
complex, the W31 H II region and SNR G11.2--0.3. 
The pixel size is $5'\times5'$.
Panels (a) and (b) display the \tqkev\ and the 1.5~keV X-ray images, 
respectively, from the mosaic of the 7 \ros PSPC pointed observations in 
Table~1 overlaid with the {\it IRAS} 100~$\mu$m contours. 
The {\it IRAS} 100~$\mu$m contours are at 500, 1000, 1500, 2000, 3000,
4500, 6000, 8000, and 10000~MJy~sr$^{-1}$.
The W30 complex and SNR G11.2--0.3 along with the detected point 
sources have been removed.  For display purpose, the data have been smoothed 
using an adaptive-filtering algorithm by selecting an area which contains 
50 counts. Thus, the resolution is variable depending on the exposure
and the total counts per pixel. The gray scale ranges from 0 to 
$700\times10^{-6}$~counts~s$^{-1}$~arcmin$^{-2}$ for panel (a) and 0 to 
$1300\times10^{-6}$~counts~s$^{-1}$~arcmin$^{-2}$ for panel (b).
\label{fig:irxr}}

\figcaption[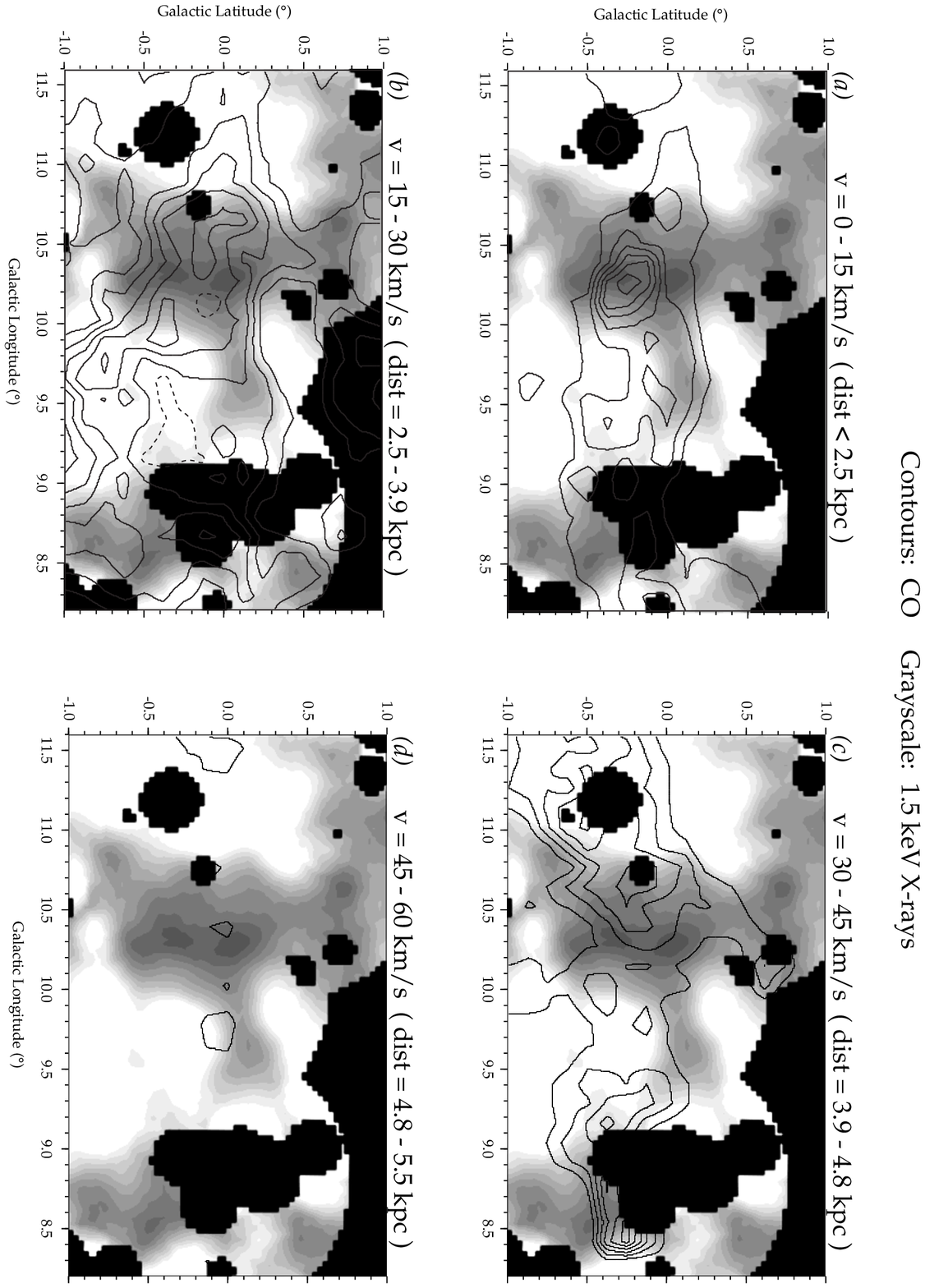]{The $^{12}$CO (J=1-0) intensity in 4 
different velocity
intervals overlaid on the gray-scale image of the 1.5 keV band X-rays 
that has been smoothed to match the 8.8$'$ angular resolution of the
CO.  The CO contours are at 28, 38, 48, 58, and 68 K~km~s$^{-1}$. The
contour at 28 K~km~s$^{-1}$ corresponds to $\sim$2 optical depths in the
$3\over4$ keV band and $\sim$1.4 optical depths in the 1.5 keV band.
The CO velocity integration range is given above each plot, along with
the corresponding range of the near kinematic distances according to the
Galactic rotation curve of Burton (1988). The X-ray gray-scale is
strongly saturated to white at high intensities in order to emphasize
the regions of absorption.
\label{fig:co}}
\end{document}